\documentclass[prd,aps,floatfix,superscriptaddress,preprint]{revtex4-1}

\pdfoutput=1
\usepackage{amsmath,amssymb}
\usepackage{bm,bbm}
\usepackage{graphicx}
\usepackage{mathrsfs}
\usepackage{slashed}
\usepackage[dvipsnames]{xcolor}
\usepackage[normalem]{ulem}
\usepackage[colorlinks=true,linkcolor=blue]{hyperref}
\usepackage{xcolor}
\usepackage{ulem}

\begin{document}

\preprint{JLAB-THY-19-3053}

\title{On the shape of the $\bar d-\bar u$ asymmetry}

\author{A.~Accardi}
\affiliation{Hampton University, Hampton, Virginia 23668, USA}
\affiliation{Jefferson Lab, Newport News, Virginia 23606, USA}
\author{C.~E.~Keppel}
\affiliation{Jefferson Lab, Newport News, Virginia 23606, USA}
\author{S.~Li}
\affiliation{University of New Hampshire, Durham, New Hampshire 03824, USA}
\author{W.~Melnitchouk}
\affiliation{Jefferson Lab, Newport News, Virginia 23606, USA}
\author{J.~F.~Owens}
\affiliation{Florida State University, Tallahassee, FL 32306, USA \\
  \vspace*{0.2cm}
  {\bf CTEQ-Jefferson Lab (CJ) Collaboration
  \vspace*{0.2cm} }}

\begin{abstract}
Using data from a recent reanalysis of neutron structure functions
extracted from inclusive proton and deuteron deep-inelastic scattering
(DIS), we re-examine the constraints on the shape of the $\bar d-\bar u$
asymmetry in the proton at large parton momentum fractions $x$.
A global analysis of the proton--neutron structure function difference
from BCDMS, NMC, SLAC and Jefferson Lab DIS measurements, and of Fermilab
Drell-Yan lepton-pair production cross sections, suggests that existing
data can be well described with $\bar d > \bar u$ for all values of $x$
currently accessible.
We compare the shape of the fitted $\bar d-\bar u$ distributions
with expectations from nonperturbative models based on chiral
symmetry breaking, which can be tested by upcoming Drell-Yan data
from the SeaQuest experiment at larger values of $x$.
\end{abstract}

\date{\today}
\maketitle

\section{Introduction}

The microscopic structure of the proton's quark--antiquark sea has
intrigued and stimulated nuclear and particle physicists for several
decades, providing a valuable window on the nonperturbative dynamics
governing quarks and gluons in QCD (see Refs.~\cite{Chang14, Speth97,
Kumano98, Vogt00, Garvey01} for reviews).
One of the most spectacular examples of how this endeavor has produced
important insights into the partonic nature of the nucleon has been
the flavor asymmetry in the light antiquark sea of the proton,
$\bar d - \bar u$.
This is expected to be negligibly small on the basis of perturbative
gluon radiation alone~\cite{Ross79}, with a scale dependence that is
suppressed by the strong coupling, $\alpha_s(Q)$.
Predicted by Thomas~\cite{Thomas83} on the basis of chiral symmetry
breaking in the strong interactions, a large excess of $\bar d$ over
$\bar u$ was, however, confirmed by several experiments involving
inclusive deep-inelastic scattering (DIS) from protons and
deuterons~\cite{NMC91, NMC94}, semi-inclusive DIS with tagging
of $\pi^+$ and $\pi^-$ mesons \cite{HERMES}, and most directly
by Drell-Yan lepton-pair production in $pp$ and $pd$ scattering
at high energies~\cite{NA51, Hawker98, Towell01}.
A~quarter of a century of experimental and theoretical efforts have
together led to a general consensus that a sizeable positive
$\bar d-\bar u$ asymmetry exists, and that its origin is likely
related to the role of the pion cloud in the nucleon, and more
generally of chiral symmetry breaking in QCD~\cite{TMS00}.

While the integrated value of the $\bar d-\bar u$ asymmetry is an
important indicator of nonperturbative physics, the shape of the
$\bar d-\bar u$ distribution itself contains even more detailed
information about the quark-gluon dynamics in the proton's sea.
In particular, the shape of the asymmetry as a function of the parton
momentum fraction, $x$, has been the source of much interest,
especially regarding its sign at large values of $x$.
Analysis of the Drell-Yan data from the Fermilab E866 experiment
\cite{Hawker98, Towell01} has suggested that the ratio of
$pd$ to $pp$ lepton-pair production cross sections drops below unity
at small values of the Feynman-$x$ variable, $x_F = 2 p_L/\sqrt{s}$,
which corresponds to large values of the partonic fraction carried
by $\bar d$ and $\bar u$ quarks in the target.
This has been interpreted as evidence for a sign change in
$\bar d-\bar u$ beyond $x \approx 0.3$, albeit within large
uncertainties, which has not been possible to accommodate in
any natural way in calculations based on chiral symmetry
breaking and the pion cloud~\cite{MST98}.

Excess of $\bar u$ over $\bar d$ was found in other approaches,
based on antisymmetrization of quark-antiquark pairs in the sea with
the valence quarks in the core of the proton.  Using a simple 3-quark
model of the nucleon with pair creation mediated by one gluon
exchange~\cite{Donoghue77}, Steffens and Thomas~\cite{Steffens97}
found that interference effects between the radiated $q\bar q$ pairs
and the core valence quarks actually generate more $u\bar u$ pairs
than $d\bar d$ pairs.
Confirmation of a sign change in the $\bar d-\bar u$ difference would
therefore be a unique signal for the presence of nonperturbative
phenomena in the nucleon sea beyond those associated with chiral
symmetry breaking.
Such effects may also be needed to explain a possibly large polarized
sea quark asymmetry $\Delta \bar d - \Delta \bar u$ in the proton
\cite{SMST92, Dressler00}, which to leading order does not receive
contributions directly from pseudoscalar meson loops.

In an interesting recent analysis, Peng {\it et~al.}
\cite{Peng-Chang-Cheng-Hou-Liu-Qiu} in fact argued that a sign change
in $\bar d-\bar u$ at intermediate $x$ is supported by an analysis of
the proton and deuteron DIS structure functions.
Combining the isovector $F_2^p-F_2^n$ structure function derived
from the NMC measurements~\cite{NMCp, NMCdop} with parametrizations
of the valence quark PDFs, Peng {\it et~al.} used a leading order (LO)
approximation for the structure functions to extract the $x$~dependence
of $\bar d-\bar u$ at $Q^2=4$~GeV$^2$, which displayed a sign change at
$x \approx 0.3$.
This intriguing behavior, along with the apparent indication of a
sign change in $\bar d-\bar u$ from the E866 data~\cite{Hawker98,
Towell01}, will soon be tested experimentally by the new SeaQuest
Drell-Yan experiment at Fermilab~\cite{SeaQuest}, which will extend
the kinematical coverage to $x \approx 0.45$.

In addition to the large-$x$ behavior, there are also questions about
the sign of $\bar d-\bar u$ at low values of $x$, below where the
current Drell-Yan data extend.
In particular, there have been indications in some global PDF analyses
for a pull to negative $\bar d - \bar u$ at low $x$, driven by the
HERA charged and neutral current DIS data~\cite{HERA}.
However, the constraining power of the HERA data for the light flavor
asymmetry at high $x$ is not as strong as the Drell-Yan data.

In this paper we revisit the question of the shape of the
$\bar d-\bar u$ asymmetry in the light of a new global analysis
of neutron structure functions~\cite{globalF2n} extracted from
inclusive proton and deuteron DIS data from experiments at
BCDMS~\cite{BCDMS}, NMC~\cite{NMCp, NMCdop}, SLAC~\cite{SLAC, SLAC-E140X}
and a new compilation of Jefferson Lab data~\cite{CJdatabase}.
Data obtained at matched kinematics --- namely, obtained from
both targets with one experimental apparatus, or within a single
experiment at the same kinematic setting --- were selected for this
analysis~\cite{JLab1, JLab2, JLab3, JLab4, JLab5, JLab6, JLab7,
JLab8, JLab9}.
Data providing ratios of the two targets, as well as a
spectator-tagged neutron structure function \cite{BONUS1, BONUS2}
measurement, were also utilized.
In particular, we compare the $F_2^p-F_2^n$ data with the structure
function difference computed self-consistently from the recent
next-to-leading order (NLO) CJ15 parton distributions~\cite{CJ15},
taking into account effects from nuclear corrections in the deuteron
and power corrections at finite $Q^2$.

We find that the existing $F_2^p-F_2^n$ data show no
evidence for a sign change in $\bar d-\bar u$ at any $x$ values,
with the zero crossing in $F_2^p-F_2^n$ entirely attributable to
NLO effects.
Furthermore, in contrast to the E866 Collaboration's extracted
$\bar d/\bar u$ ratio, the $pd$ to $pp$ Drell-Yan cross section
ratio is well described in terms of the CJ15 PDFs, for which
$\bar d > \bar u$ at all values of $x$.
Finally, we compare the shape of the $\bar d-\bar u$ asymmetry with
expectations from nonperturbative models of the nucleon based on
chiral symmetry breaking, and stress the need for consistent,
global QCD analysis of all data before robust conclusions about
the shape and sign of $\bar d-\bar u$ can be drawn.

\section{Isovector nucleon structure function}

As observed by Peng {\it et~al.}~\cite{Peng-Chang-Cheng-Hou-Liu-Qiu},
if one writes the proton and neutron $F_2$ structure functions at LO
in terms of PDFs, then the difference $\bar d-\bar u$ can be obtained
from the isovector $F_2^p-F_2^n$ structure function combination and the
difference between the $u$ and $d$ valence quark PDFs in the proton,
\begin{equation}
\label{eq.Peng}
\Delta(x)\,
\equiv\,  \frac12 \Big[ u_v(x) - d_v(x) \Big]
	- \frac{3}{2x} \Big[ F_2^p(x) - F_2^n(x) \Big].
\end{equation}
At LO, one obviously has $\Delta(x) = \bar d(x) - \bar u(x)$.
At higher orders, the quantity defined in (\ref{eq.Peng}) will not be
identical to $\bar d(x) - \bar u(x)$.
In their analysis, Peng {\it et~al.} proceed to extract $\Delta(x)$
from the $F_2^p-F_2^n$ difference derived from the NMC data~\cite{NMCp,
NMCdop} by combining this with the valence PDFs obtained from the
JR14~\cite{JR14} and CT10~\cite{CT10} parametrizations at NNLO.
The result was found to produce a sign change at $x \sim 0.3$,
which was interpreted as a zero crossing of $\bar d(x)-\bar u(x)$.

Peng~{\it et~al.} argue~\cite{Peng-Chang-Cheng-Hou-Liu-Qiu} that since
the integrated value of $\bar d-\bar u$, and the associated Gottfried
sum~\cite{GSR}, receive very small ${\cal O}(\alpha_s)$~\cite{Ross79}
and ${\cal O}(\alpha_s^2)$~\cite{Kataev03} corrections, the LO
approximation~(\ref{eq.Peng}) should be accurate.
However, while the correction to the integrated value of
$\bar d-\bar u$ is indeed small~\cite{Ross79}, the higher order
effects on the $x$~dependence of the asymmetry may not be negligible.
This could in practice then lead to misidentification of perturbative
higher order effects with the behavior of the nonperturbative parton 
distributions as a function of $x$, as we discuss in the following.

\begin{figure}[t]
\begin{center}
\hspace*{1.5cm}\includegraphics[width=1.4\textwidth]{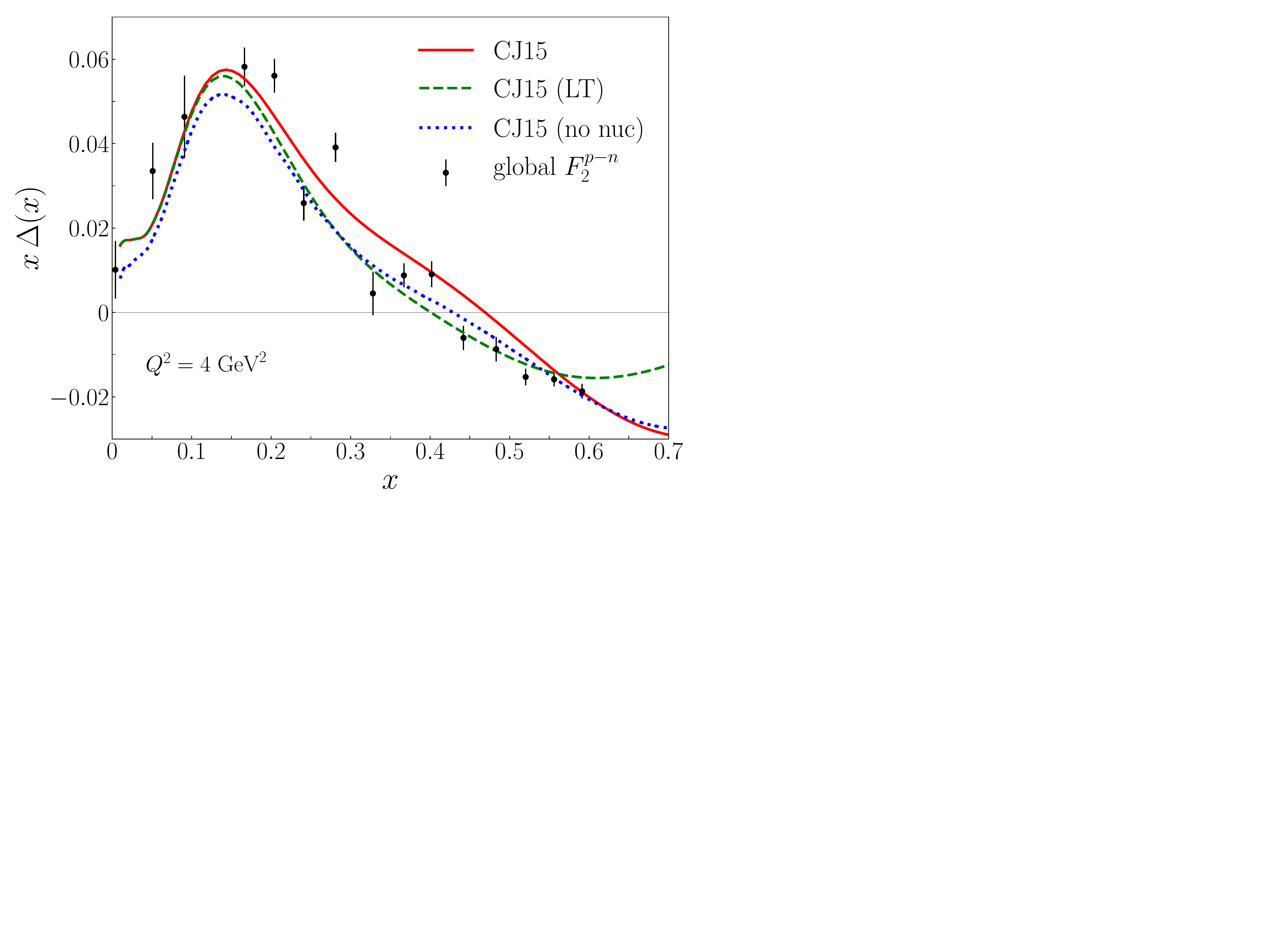}
\vspace*{-9cm}
\caption{Isovector combination $x \Delta$, defined in Eq.~(\ref{eq.Peng}),
	computed from the CJ15 NLO PDFs \cite{CJ15} (red solid curve)
	at $Q^2=4$~GeV$^2$, and compared with $\Delta$ calculated in the
	leading twist approximation (green dashed curve), and neglecting
	nuclear corrections in the deuteron (blue dotted curve).
	The data points (black circles) are from the global neutron
	structure function analysis~\cite{globalF2n, Niculescu}
	using the CJ15 valence quark~PDFs.}
\label{fig.Delta_DIS}
\end{center}
\end{figure}

To quantify this effect, we compute the quantity $\Delta(x)$ in
Eq.~(\ref{eq.Peng}) using the CJ15 NLO parton distributions~\cite{CJ15}
for all terms on the right hand side of the equation.
This is shown in Fig.~\ref{fig.Delta_DIS} at $Q^2=4$~GeV$^2$, where
the calculated $\Delta(x)$ is compared with the corresponding quantity
constructed from the global $F_2^p-F_2^n$ data~\cite{globalF2n},
using with the CJ15 parametrization for the valence $u_v-d_v$ PDFs.
Both the calculated $\Delta(x)$ and the result extracted from the
global data peak at $x \sim 0.1-0.2$, before decreasing at higher $x$
and turning negative at $x \gtrsim 0.4$.
The general agreement between the calculated and phenomenological
$\Delta$ results suggests that the CJ15 fit is able to describe well
the global $F_2^p-F_2^n$ data, including the change in sign at large $x$.

This remains the case irrespective of finite-$Q^2$ power corrections
or nuclear effects, as Fig.~\ref{fig.Delta_DIS} illustrates.
In particular, since the value of $Q^2=4$~GeV$^2$ is not particularly
high, one could imagine that finite-$Q^2$ corrections, associated with
target mass effects or higher twists \cite{Schienbein08, Moffat19},
may impact the shape of $\Delta(x)$.
To examine this possibility we compute the $F_2^p-F_2^n$ structure
function difference in Eq.~(\ref{eq.Peng}) from the CJ15 PDFs at
leading twist (LT) only, without the finite-$Q^2$ corrections.
Comparison with the full result in Fig.~\ref{fig.Delta_DIS} shows
that the result is only slightly modified by the finite-$Q^2$ effects,
with the zero crossing of $\Delta$ at $x \approx 0.4$ remaining.

A further complication in the application of Eq.~(\ref{eq.Peng})
arises from the possible nuclear effects that may obscure the
extraction of the neutron $F_2^n$ structure function from the
inclusive proton and deuteron DIS data.
In the CJ15 global analysis the nuclear effects in the deuteron
were taken into account through a systematic expansion in the weak
binding approximation~\cite{KP06, KMK09}, in which nuclear binding
and Fermi motion effects are described through nucleon smearing
functions, and nucleon off-shell corrections~\cite{KP06, MSToff,
MSTplb, MTnp} are parametrized phenomenologically.
To quantify the nuclear effect we therefore compute $\Delta$ from
the CJ15 PDFs, but with the $F_2^n$ calculated as the difference
between the deuteron and proton structure functions, without any
nuclear corrections, $F_2^n = F_2^d - F_2^p$.
Again, we see no qualitative difference between the uncorrected
and nuclear corrected neutron structure function.

\begin{figure}[t]
\begin{center}
\hspace*{1.5cm}\includegraphics[width=1.4\textwidth]{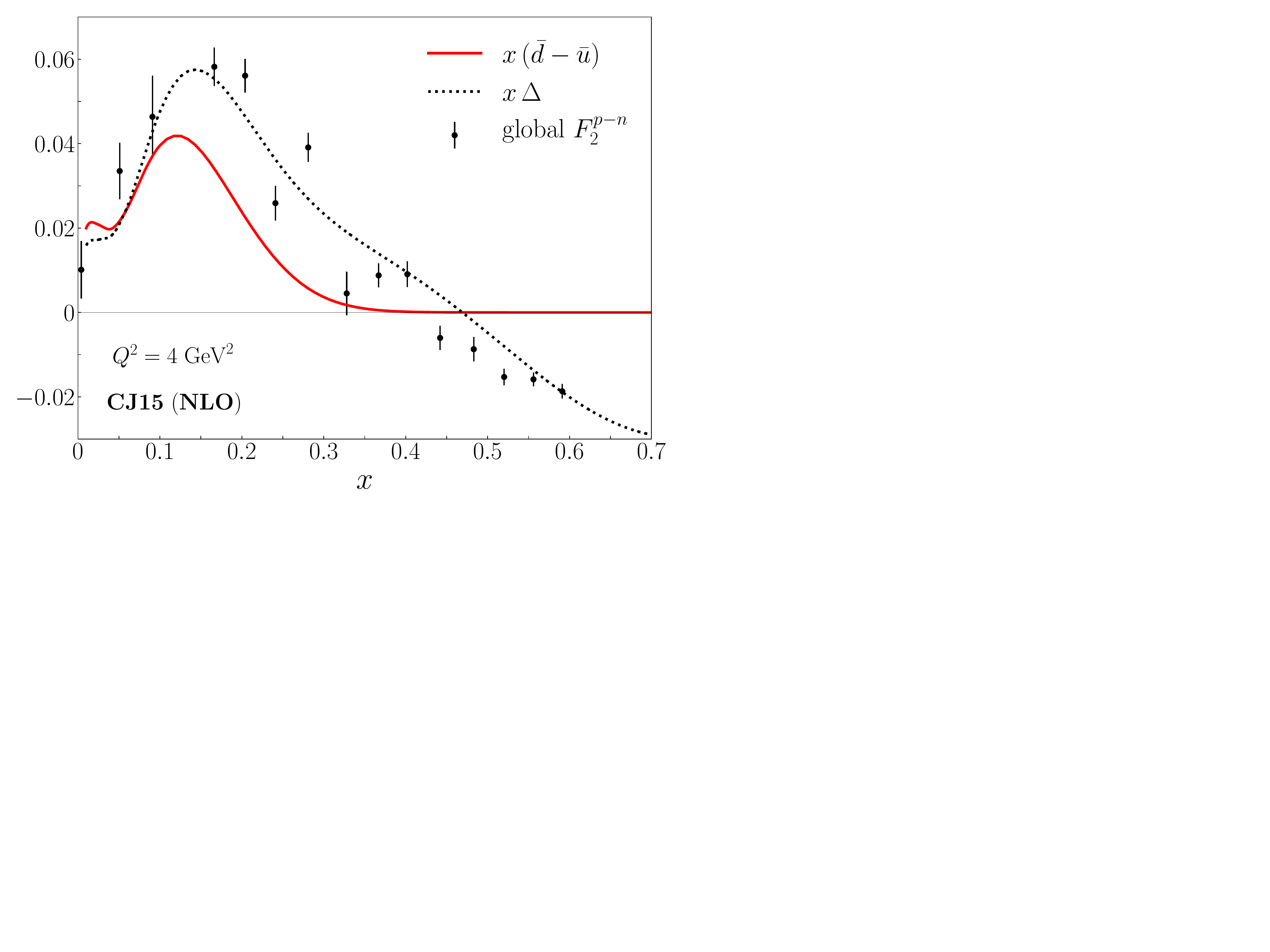}
\vspace*{-9cm}
\caption{Antiquark asymmetry $x(\bar d-\bar u)$ from the CJ15 global
	QCD analysis~\cite{CJ15} at $Q^2=4$~GeV$^2$ (red solid curve)
	compared with the phenomenological $x \Delta$ constructed from
	the proton--neutron structure function data~\cite{globalF2n,
	Niculescu} (black circles), and with $\Delta$ computed from
	the CJ15 fit (black dotted curve).}
\label{fig.dbub_DIS}
\end{center}
\end{figure}

While the $x$ dependence of the phenomenological $\Delta$ is
consistent with the calculation based on the CJ15 NLO PDFs~\cite{CJ15},
we should note that the same global QCD analysis has, by construction,
a $\bar d-\bar u$ asymmetry that is positive definite for all $x$,
as illustrated in Fig.~\ref{fig.dbub_DIS}.
In particular, while at LO the quantities $\Delta$ and $\bar d-\bar u$
coincide, at NLO or at higher order there is no reason for a sign
change in $\Delta$ to require a sign change in $\bar d-\bar u$.
A negative $\Delta$ is naturally generated by higher order $\alpha_s$
effects and other corrections that significantly modify the shape of
the $x$~dependence at intermediate and large values of $x$.

The comparisons in Figs.~\ref{fig.Delta_DIS} and \ref{fig.dbub_DIS}
plainly demonstrate that the apparent sign change in the $\bar d-\bar u$
difference extracted from $F_2^p-F_2^n$ is indeed an artifact induced
by higher order QCD corrections, which affect in a nontrivial way the
shape of the $x$ distribution of the structure functions.
On the other hand, it has long been accepted that the Fermilab E866
Drell-Yan data clearly indicate that the $\bar d/\bar u$ ratio,
extracted from the ratio of $pd$ to $pp$ lepton-pair production
cross sections, drops below unity at $x \gtrsim 0.3$~\cite{Hawker98,
Towell01}.
We discuss the Drell-Yan data and their implications in more detail
next.

\section{Drell-Yan cross sections}

The strongest evidence for a nonzero $\bar d-\bar u$ asymmetry has come
from the Fermilab E866 Drell-Yan experiment~\cite{Hawker98, Towell01},
which measured the ratio of $pd$ to $pp$ lepton-pair production cross
sections at an average $Q^2=54$~GeV$^2$.
At LO, the cross section is proportional to a sum over flavors $q$ of
products of PDFs in the beam ($b$) and target ($t$) hadrons, evaluated
at parton momentum fractions $x_b$ and $x_t$, respectively~\cite{DY70},
\begin{equation}
\frac{d\sigma}{dx_F dQ^2}\
\propto\ \sum_q\, e_q^2\, 
    \big[ q_b(x_b)\, \bar q_t(x_t) + \bar q_b(x_b)\, q_t(x_t)
    \big],
\label{eq:dy}
\end{equation}
where
    $x_F = x_b - x_t$
is the Feynman scaling variable, and
    $x_b\, x_t \approx Q^2/s$,
with $Q$ the invariant mass of the dilepton pair, and $\sqrt{s} \approx 
40$~GeV is the center of mass energy at the E866 kinematics.
Furthermore, for $x_b \gg x_t$ the cross section ratio at LO simplifies 
to a ratio that depends only on the antiquark PDFs in the 
target~\cite{Ellis91},
\begin{equation}
\label{eq.DYrat}
\frac{\sigma^{pd}}{\sigma^{pp}}
\approx 1 + \frac{\bar d(x_t)}{\bar u(x_t)}.
\end{equation}
In practice, the E866/NuSea Collaboration extracted the $\bar d/\bar u$ 
ratio using an iterative procedure to take into account experimental 
acceptance corrections, assuming that existing PDF parametrizations at 
the time~\cite{MRST98, CTEQ5} accurately described the valence and 
heavy-quark distributions, as well as the sum, $\bar d+\bar u$, of the 
light antiquark PDFs~\cite{Towell01}.
From the $\bar d/\bar u$ ratio the difference $\bar d-\bar u$ was then 
computed at the E866 kinematics assuming $\bar d+\bar u$ from 
Ref.~\cite{CTEQ5}.

\begin{figure}[t]
\hspace*{1.5cm}\includegraphics[width=1.4\textwidth]{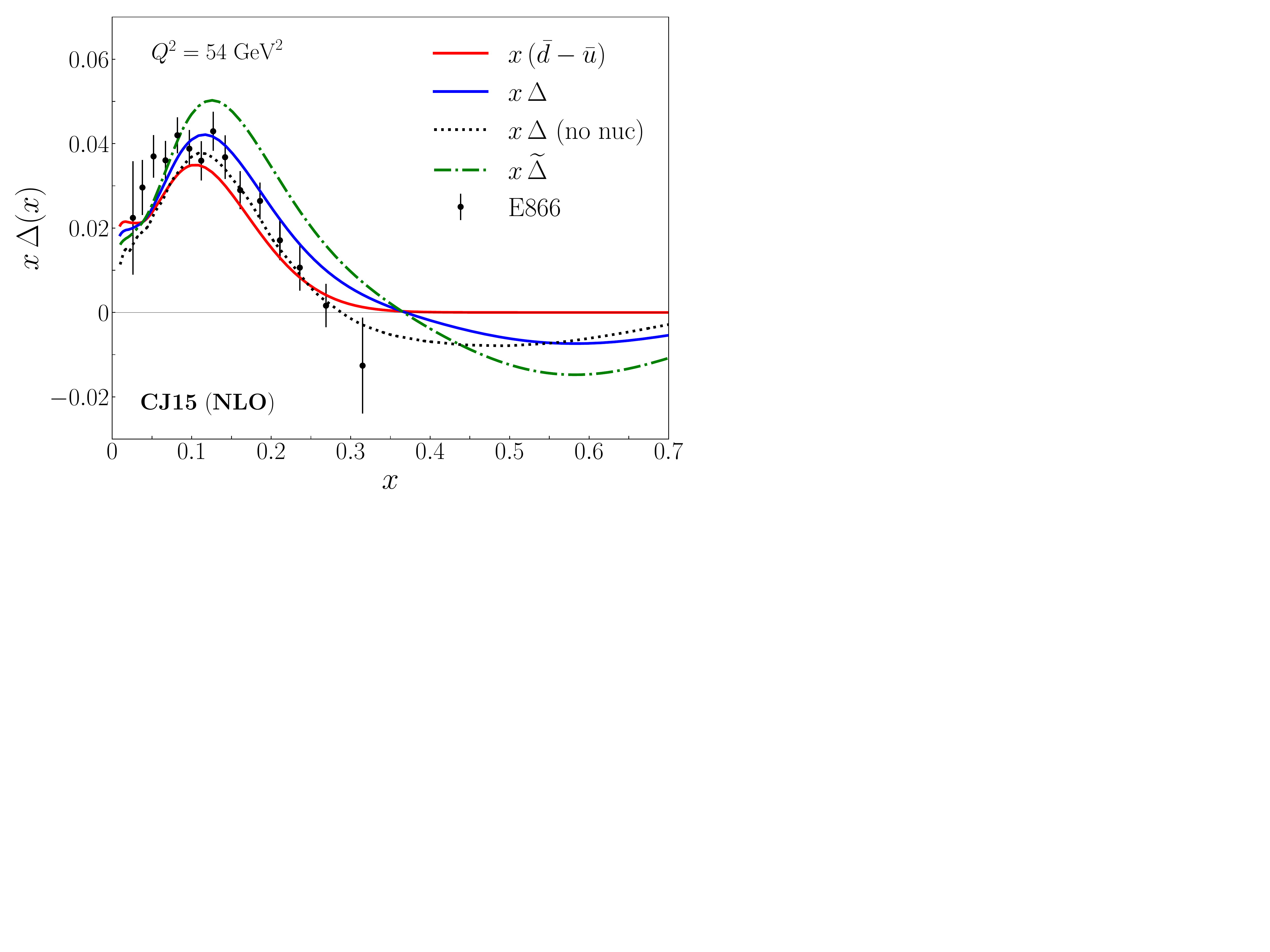}
\vspace*{-9cm}
\caption{Antiquark asymmetry $x (\bar d-\bar u)$ from the CJ15 NLO
        parametrization~\cite{CJ15} (red solid curve) at $Q^2=54$~GeV$^2$
        compared with the values extracted from the ratio of $pd$ to $pp$
        Drell-Yan cross sections, assuming $\bar d+\bar u$ from
        Ref.~\cite{CTEQ5} (black circles), and with the isovector
        combination $x \Delta$ defined in Eq.~(\ref{eq.Peng})
        computed from the CJ15 PDFs (blue solid curve),
        with $\Delta$ computed neglecting nuclear effects in the
        deuteron (black dotted curve), and with an alternative
        definition in Eq.~(\ref{eq.Peng2}) (green dot-dashed curve)
        at the same $Q^2$ value.}
\label{fig.Delta_DY}
\end{figure}

The resulting $\bar d-\bar u$ values are shown in 
Fig.~\ref{fig.Delta_DY} at the average $Q^2=54$~GeV$^2$, illustrating 
the strong enhancement of the asymmetry at $x \approx 0.1$, and the 
tendency towards negative values for $x \gtrsim 0.3$.
The latter trend is similar to that displayed by the isovector 
combination $\Delta$, computed from the CJ15 NLO PDFs~\cite{CJ15} with 
or without nuclear effects in the deuteron, as in 
Fig.~\ref{fig.Delta_DIS}.
On the other hand, the actual $\bar d-\bar u$ difference from the CJ15 
parametrization at the same $Q^2$ remains positive definite at all $x$ 
values, as in the comparison with the DIS data in 
Fig.~\ref{fig.Delta_DIS} at the lower $Q^2$.

In fact, the relation (\ref{eq.Peng}) for the isovector distribution
$\Delta$, used as the basis for the analysis in 
Ref.~\cite{Peng-Chang-Cheng-Hou-Liu-Qiu}, is not the only
representation of the sea asymmetry.
An alternative representation, which is equivalent to 
Eq.~(\ref{eq.Peng}) at LO, relates $\bar d-\bar u$ to the isovector 
structure function $F_2^p-F_2^n$ and the total $u$ and $d$ quark PDF 
difference, rather than to the $u_v-d_v$ valence distributions,
\begin{equation}
\label{eq.Peng2}
\widetilde\Delta(x)
\equiv u(x) - d(x) - \frac{3}{x} \Big[ F_2^p(x) - F_2^n(x) \Big].
\end{equation}
At LO in $\alpha_s$, obviously
	$\Delta = \widetilde\Delta = \bar d - \bar u$;
however, at higher orders Eqs.~(\ref{eq.Peng}) and (\ref{eq.Peng2})
are not identical.
The differences between $\Delta$ and $\widetilde\Delta$ at
$Q^2=54$~GeV$^2$ are shown in Fig.~\ref{fig.Delta_DY}, and
reveal discrepancies of $\sim 20\%-30\%$ at $x \sim 0.1-0.3$,
and even greater at larger $x$ values, $x \sim 0.5$.
Of course, other definitions for the isovector combination
$\Delta$ could also be used, which all have the same LO limit,
but introduce arbitrary differences at higher orders.

\begin{figure}[t]
\hspace*{1.5cm}\includegraphics[width=1.4\textwidth]{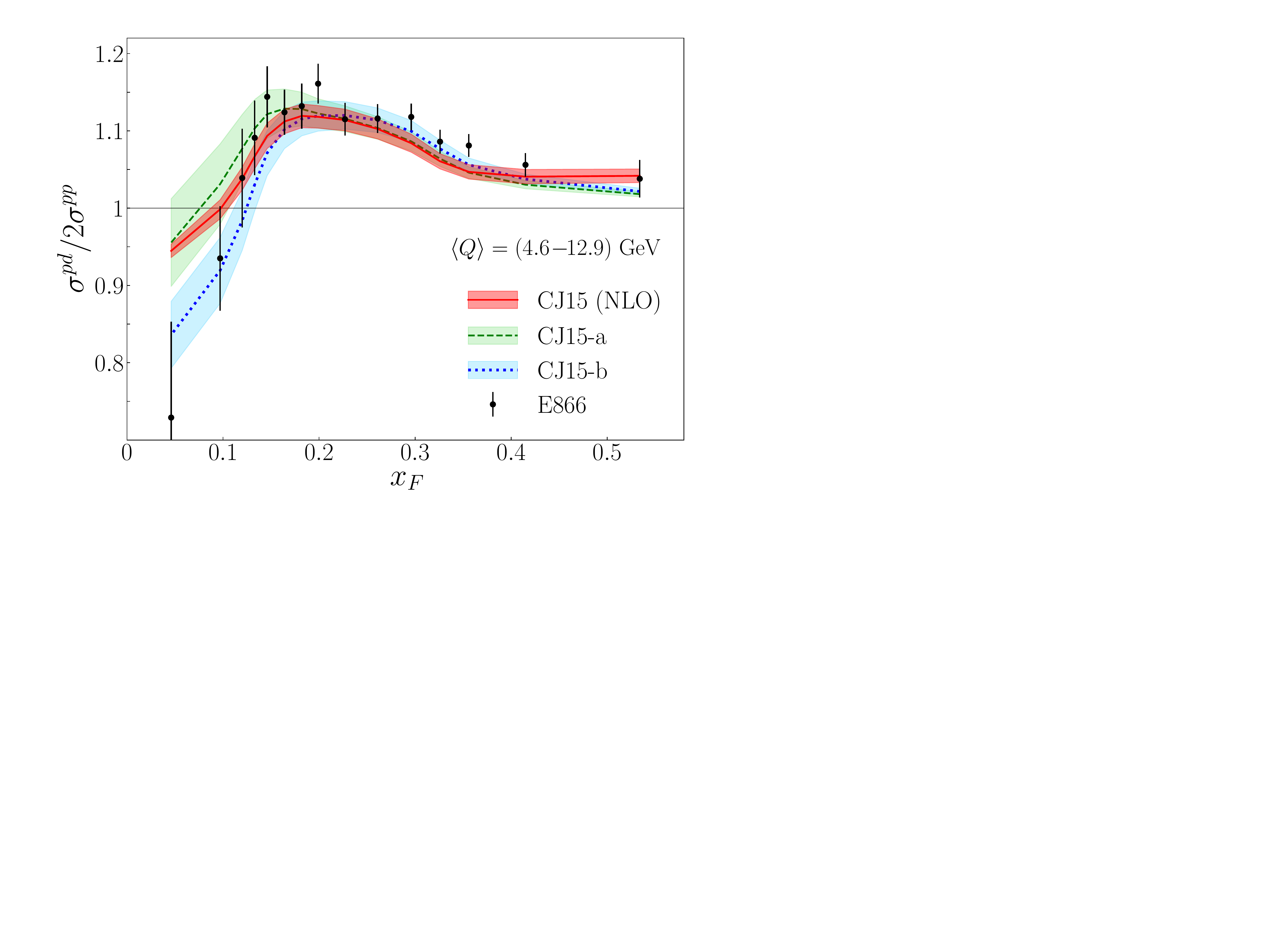}
\vspace*{-9cm}
\caption{Ratio of Drell-Yan lepton-pair production cross sections for
    	$pd$ and $pp$ collisions from the Fermilab E866 experiment
	\cite{Towell01} versus the Feynman variable $x_F$, compared
        with the ratios calculated from the CJ15~\cite{CJ15} PDFs
        (red solid curve and band) and from a variation (CJ15-a)
        of the fit which parametrizes the difference $\bar d-\bar u$
        instead of the ratio (green dashed curve and band),
        and a fit (CJ15-b) using data as in the CJ12~\cite{CJ12}
        analysis (blue dotted curve and band).
    	The average values of $Q$ range from
    	4.6~GeV (at the highest	$x_F$) to
    	12.9~GeV (at the lowest~$x_F$).}
\label{fig.sigratio}
\end{figure}

This illustrates the intrinsic ambiguities inherent in comparing 
quantities extracted from cross sections with inconsistent use of 
perturbative QCD corrections.
The most robust and unambiguous way to compare experimental data with 
theory is to directly compute the observables in terms of PDFs at a 
given order in $\alpha_s$, using universal PDFs extracted from other 
data sets at the same order, as is typically done in global QCD 
analyses~\cite{JMO13, Forte13, PDF4LHC16}.
We highlight this in Fig.~\ref{fig.sigratio}, which shows the actual 
experimentally measured ratio of $pd$ to $pp$ Drell-Yan cross sections 
from the E866 experiment versus the Feynman variable $x_F$, with the 
average $Q$ ranging from 4.6~GeV at the highest $x_F$ to 12.9~GeV at
the lowest $x_F$ points.
From the kinematics of the Drell-Yan process, high $x_F$ values 
correspond to low $x_t$ values, and the lowest $x_F$ correspond to
the highest $x_t$, which are most sensitive to the $\bar d/\bar u$
ratio in the target hadron.

The ratio computed from the CJ15 PDFs is generally in good agreement 
with the measured ratio across all $x_F$.
Note that the CJ15 analysis fitted the absolute $pp$ and $pd$ Drell-Yan 
cross sections, rather than the derived cross section ratio, giving an 
overall
    $\chi^2$ per datum of $284/250 \approx 1.14$,
using a cut on dimuon masses of $Q > 6$~GeV~\cite{CJ15}.
As illustrated in Figs.~\ref{fig.Delta_DIS} and~\ref{fig.Delta_DY}, a 
$\bar d-\bar u$ difference that is always positive (or, equivalently, 
$\bar d/\bar u$ ratio always above unity) can nonetheless give rise to 
observables (structure functions or cross sections) that naively would 
suggest a sign change at LO.
The dip below unity of the Drell-Yan cross section ratio evident at
low $x_F$, $x_F \lesssim 0.1$, in Fig.~\ref{fig.Delta_DIS} is an
example of this.

\begin{figure}[t]
\begin{center}
\hspace*{1.5cm}\includegraphics[width=1.4\textwidth]{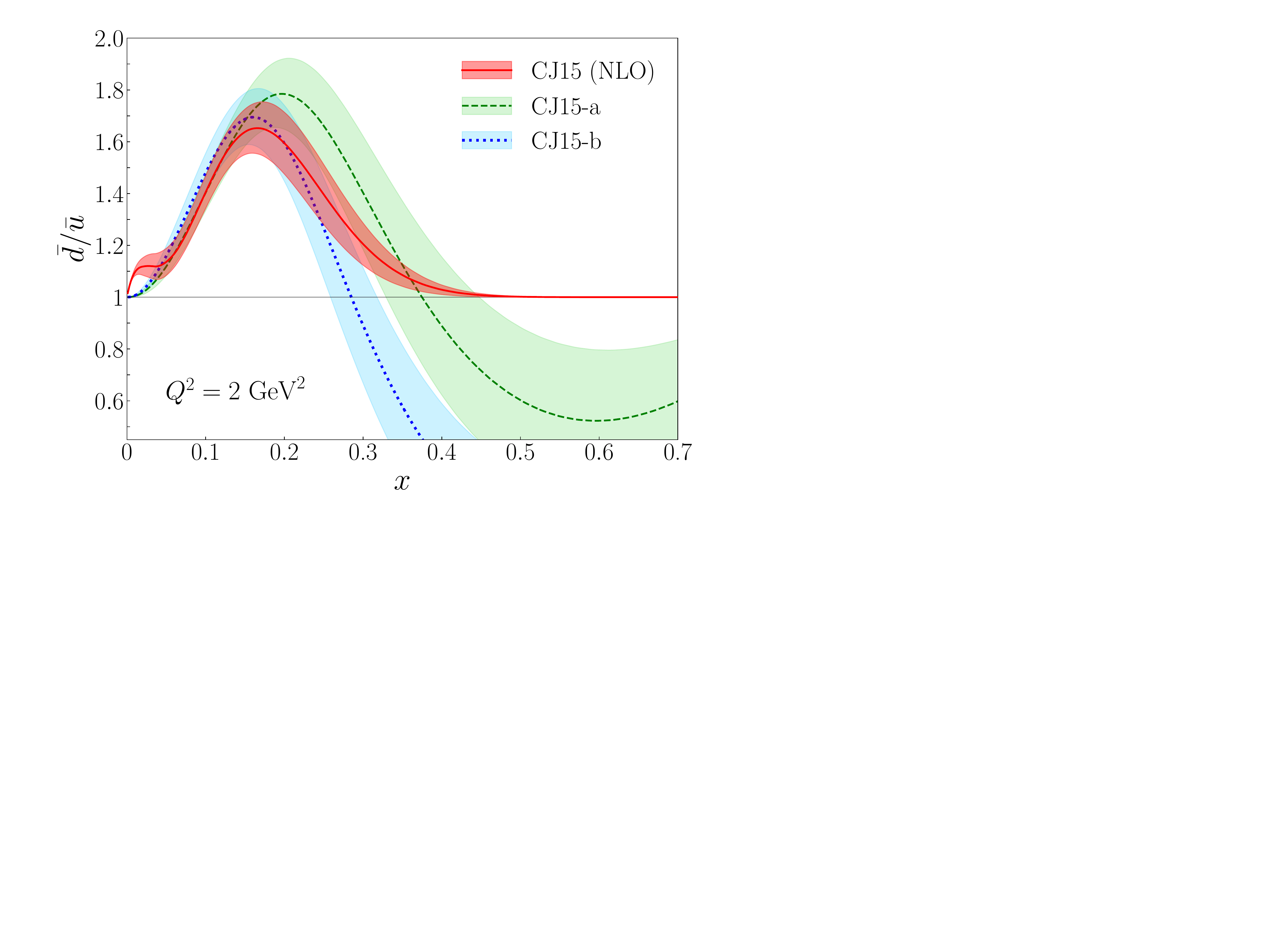}
\vspace*{-9cm}
\caption{Ratio of $\bar d/\bar u$ PDFs at a scale $Q^2=2$~GeV$^2$
	from the CJ15~\cite{CJ15} NLO parameterizations
	(red solid curve and band),
        compared with the ratio from the variant
        CJ15-a (green dashed curve and band) and
        CJ15-b (blue dotted curve and band) fits,
        both of which dip below 1 at large values of~$x$.}
\label{fig.dbub}
\end{center}
\end{figure}

In fact, a similar behavior is also found if one replaces the 
positive-definite parametrization of $\bar d/\bar u$ used in the CJ15 
fit with the more conventional parametrization of the difference,
    $(\bar d - \bar u)(x)
	= N x^\alpha(1-x^\beta)
	  (1 + \gamma \sqrt x + \delta x)$,
as employed for example in the earlier CJ12 analysis~\cite{CJ12}.
This parametrization then allows the $\bar d$ PDF to be smaller than the 
$\bar u$ in some regions of $x$.
The resulting fit, however, which we denote by ``CJ15-a'', also 
reproduces the E866 cross section ratio quite well, as 
Fig.~\ref{fig.sigratio} illustrates, with a similar
    $\chi^2$ per datum of $294/250 \approx 1.18$.
Interestingly, the $\bar d/\bar u$ ratio in the CJ15-a fit remains
above unity up to parton momentum fractions $x \approx 0.4$,
and is even slightly higher than in the standard CJ15 fit,
as Fig.~\ref{fig.dbub} illustrates, before dipping below 1 at
$x \gtrsim 0.4$.
This shows that the positivity of the antiquark ratio is driven
by data and is not an artifact of the chosen parametrization.
Note that with more parameters in the CJ15-a parametrization,
the resulting error band on the $\bar d/\bar u$ ratio is larger.
Conversely, the standard CJ15 parametrization is less flexible and is 
therefore more tightly constrained by the data, with the resulting 
uncertainty band being smaller.

In order to examine the effect on the $\bar d - \bar u$ shape at large 
$x$ from the interplay between the choice of parametrization and the 
data sets used in the global analysis, we perform a further fit in which 
the CJ15 data sets are replaced by the data that were used in the CJ12 
analysis~\cite{CJ12}, while retaining the QCD theory setup as in 
CJ15~\cite{CJ15}, as well as the more flexible parametrization utilized 
for CJ15-a.
We refer to this fit as ``CJ15-b''.
As far as the impact on the antiquark PDFs, the main difference between 
the data sets utilized in the CJ15 and CJ15-a analyses compared to 
CJ15-b are the more stringent cut on the dilepton mass of $Q>6$~GeV in 
the E866 Drell-Yan data~\cite{Towell01} and the use of newer $W$-boson 
charge asymmetry data from D0~\cite{D0_13, D0_15}.
The more relaxed cut of $Q>4$~GeV in CJ15-b increases the number of 
available data points by $\sim 50\%$, allowing better constraints on the 
low-$x_F$ cross section ratio, as evident in Fig.~\ref{fig.sigratio}.
This is achieved through the generation of a stronger dip in the $\bar 
d/\bar u$ ratio below unity at $x \gtrsim 0.3$, as illustrated in 
Fig.~\ref{fig.dbub}.
However, the overall fit to the E866 cross sections across all 
kinematics becomes somewhat worse, with a
    $\chi^2$ per datum of $593/375 \approx 1.58$.
This is mostly due to the difficulty in fitting the $pd$ cross section 
data at low-$Q$ values, which were shown to be in tension with fixed 
target DIS data~\cite{Alekhin06}.
When the earlier, less precise D0 $W$ asymmetry data are replaced by
the more recent and more precise results~\cite{D0_13, D0_15}, the dip
is reduced significantly.

Note also that for the E866 $pd$ data, the lowest $x_F$ kinematics 
involve deuteron parton momentum fractions $x_t \approx 0.25-0.35$, at 
which Fermi smearing and binding effects may start to become relevant. 
Ehlers {\it et~al.}~\cite{Ehlers14} studied these effects quantitatively 
within the same framework as used for DIS from the deuteron~\cite{KP06, 
KMK09}, including the possible off-shell modifications of the nucleon 
PDFs in medium~\cite{MSToff, MSTplb}.
While increasing in strength at higher $x$ values, where there is 
greater sensitivity to the large momentum components of the deuteron 
wave function, the nuclear effects were found to be relatively small on 
the scale of the uncertainties on the E866 cross section ratio data.
However, the nuclear corrections will become more important at the 
higher $x$ values of the new SeaQuest experiment~\cite{SeaQuest}, 
especially with the expected reduction in experimental uncertainties.

\section{Outlook}

With the SeaQuest Drell-Yan data anticipated in the very near future, 
the kinematic coverage over which the $\bar d-\bar u$ difference
can be directly constrained is expected to extend to
$x \approx 0.45$~\cite{SeaQuest}.
In particular, in the region $x \approx 0.25-0.3$, where the E866 
data~\cite{Towell01} suggested a possible cross-over of the
$\bar d/\bar u$ ratio, the experimental uncertainties on the
new measurements should be sufficiently small to verify whether
this is indeed a robust feature of the high-$x$ data.
This should allow more definitive conclusions to be reached about the 
sign of the $\bar d-\bar u$ difference, and whether chiral symmetry 
breaking considerations alone can account for the shape of the 
asymmetry~\cite{MST98, Salamu15, NonlocalI, NonlocalII} or additional 
physical mechanisms are needed~\cite{Donoghue77, Steffens97, SMST92, 
StatPDF, StatPDF-2, Field77}.

\begin{figure}[t]
\begin{center}
\includegraphics[width=0.7\textwidth]{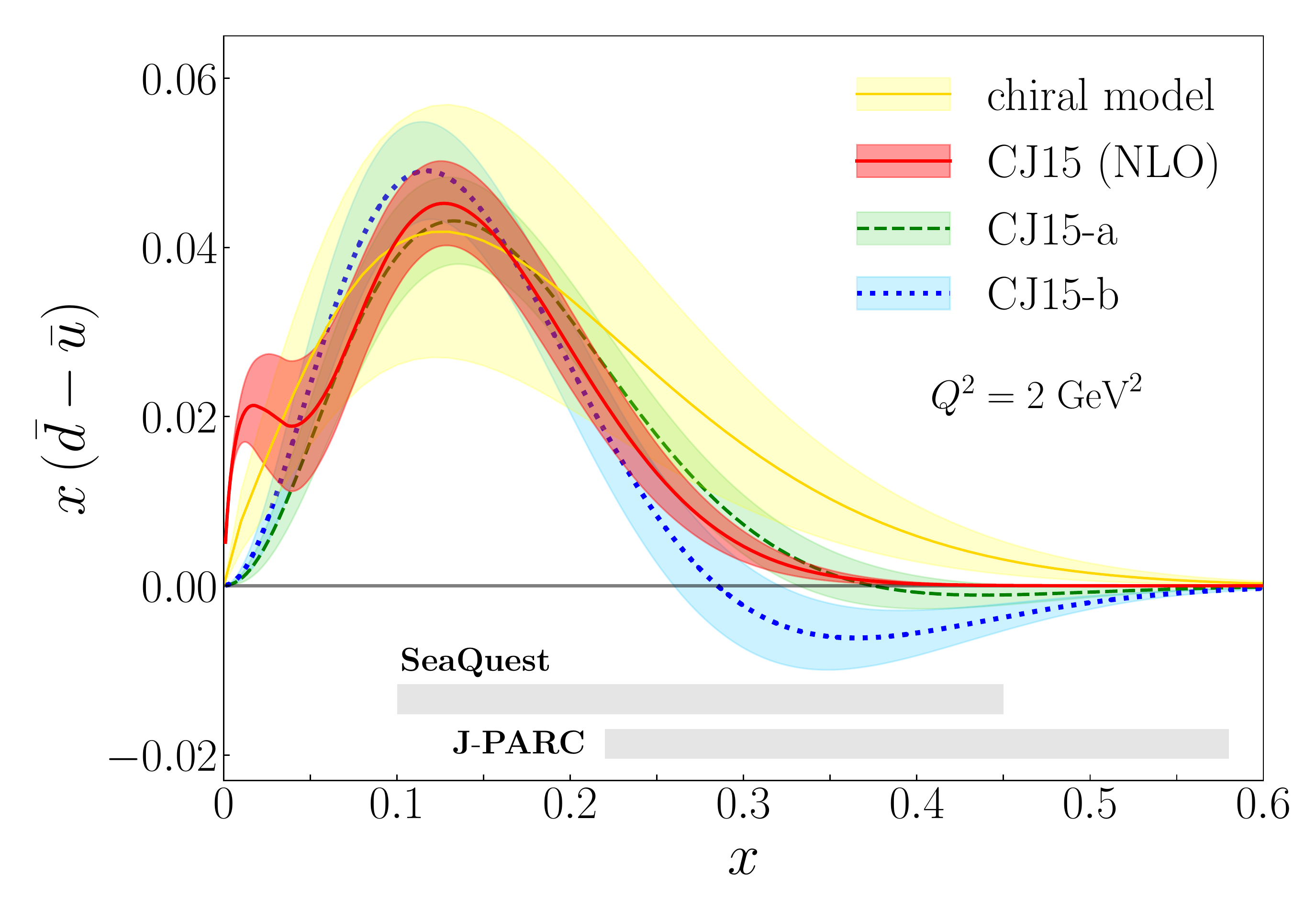}
\caption{Momentum dependence of the antiquark asymmetry
	$x(\bar d-\bar u)$ at a scale $Q^2=2$~GeV$^2$ from the
	CJ15~\cite{CJ15} NLO fit (red dashed curve and band)
	and the CJ15-a (green dashed curve and band) and
	CJ15-b (blue dotted curve and band) variations,
	compared with a nonperturbative calculation of
	pion loop contributions from chiral effective
	theory~\cite{Salamu15, NonlocalI, NonlocalII}.
        The expected kinematical coverage of the future
        SeaQuest~\cite{SeaQuest} and J-PARC~\cite{J-PARC}
        experiments is indicated by the horizontal
        gray bands.}
\label{fig.xdbub}
\end{center}
\end{figure}

As Fig.~\ref{fig.xdbub} demonstrates, precise data will be needed
to discriminate between the different possible behaviors of the
$\bar d-\bar u$ asymmetry at $x \gtrsim 0.2$.
While all 3 analyses considered here (the standard CJ15 and the two 
variants, CJ15-a and CJ15-b) produce results for $x(\bar d-\bar u)$ 
which display strong positive peaks at $x \approx 0.1$, the modified 
CJ15-b fit drops faster and crosses zero at $x \approx 0.25-0.3$, 
whereas the asymmetry in the standard CJ15 fit remains positive.
As illustrated in Fig.~\ref{fig.Delta_DY}, all 3 variants give good 
descriptions of the E866 Drell-Yan data, with equally good $\chi^2$ 
values, and the differences between the sets of distributions reflect 
the limitations of existing data in constraining the high-$x$ behavior.
The differences between these parametrizations is also fairly
indicative of the spread in $\bar d-\bar u$ from other global QCD
analyses~\cite{CT14, MMHT14, NNPDF3.1, ABMP16, JAM19} that use the
E866 data.

Upcoming data from the Fermilab SeaQuest experiment~\cite{SeaQuest},
as well as future data from the proposed Drell-Yan experiment at
J-PARC~\cite{J-PARC}, will constrain the $\bar d-\bar u$ asymmetry
out to $x \approx 0.45$ and $\approx 0.55-0.6$, respectively.
%
%
With sufficient precision, the new data should help answer the question
whether $\bar d-\bar u$ changes sign or stays positive at high $x$,
as predicted in models based on chiral symmetry breaking.
In particular, the latter involve convolution of PDFs in the pion and
splitting functions for the proton to baryon plus pion conversion.
The hadronic splitting is dominated by the (positive) contributions
from the $p \to n \pi^+$ process, with smaller (negative) contributions
from the $p \to \Delta^0 \pi^-$ dissociation.
Phenomenologically, it is very difficult to accommodate a negative
overall contribution to $\bar d-\bar u$ at any value 
of~$x$~\cite{MST98}, and a typical result for the asymmetry from
chiral loops is illustrated in Fig.~\ref{fig.xdbub} from
Ref.~\cite{NonlocalII}.

\newpage
Of course, additional mechanisms beyond those associated with chiral 
symmetry breaking, such as those based on the Pauli exclusion 
principle~\cite{Donoghue77, Steffens97, SMST92, StatPDF, StatPDF-2, 
Field77}, may play a role in generating some of the asymmetry.
Whether and to what extent such mechanisms are important
phenomenologically may be revealed with the upcoming Drell-Yan
data~\cite{SeaQuest, J-PARC}.

\acknowledgements

We thank M.~Diefenthaler, G.~Niculescu, I.~Niculescu, J.-C.~Peng and
J.~Qiu for helpful discussions and communications.
This work was supported by the U.S. Department of Energy (DOE)
Contract No.~DE-AC05-06OR23177, under which Jefferson Science
Associates, LLC operates Jefferson Lab. \\


\end{document}